\documentclass[12pt]{article}
\usepackage{scicite}
\usepackage{times}
\usepackage[dvipsnames]{xcolor}
\usepackage{graphicx}
\usepackage{bbold}
\usepackage[normalem]{ulem}

\usepackage{physics}
\usepackage{dcolumn}
\usepackage{bm}
\usepackage{amssymb}
\usepackage{subfigure}
\usepackage{here}
\usepackage{changes}
\usepackage{enumerate}
\usepackage{cancel}
\usepackage{flushend}
\usepackage{bbold}
\usepackage{mathtools}
\usepackage{float}
\usepackage{stmaryrd}
\usepackage{colortbl}
\usepackage{amsmath,amssymb,amsfonts,stmaryrd,wasysym,graphicx,multirow,color,textcomp,subfigure}
\usepackage{url}
\usepackage{romannum}
\usepackage[colorlinks=true,citecolor=blue,urlcolor=blue]{hyperref}
\newcommand{\eq}[1]{Eq.~(\ref{#1})}

\newcommand{\quoting}[1]{``#1''}

\def\be{\begin{equation}}
	\def\ee{\end{equation}}
\def\bea{\begin{eqnarray}}
	\def\eea{\end{eqnarray}}

\makeatletter
\newcommand*\bigcdot{\mathpalette\bigcdot@{.5}}
\newcommand*\bigcdot@[2]{\mathbin{\vcenter{\hbox{\scalebox{#2}{$\m@th#1\bullet$}}}}}
\makeatother

\newenvironment{sciabstract}{%
\begin{quote} \bf}
{\end{quote}}

\title{Bloch Theorem Dictated Wave Chaos\\ in Microcavity Crystals} 

\author
{Chang-Hwan Yi$^1$, Hee Chul Park$^{1,2,a \dagger}$, Moon Jip Park$^{1,3,b \dagger}$\vspace{-0.1cm}\\
\\
\normalsize{$^1$Center for Theoretical Physics of Complex Systems, Institute for Basic Science (IBS), }\\
\normalsize{Daejeon, 34126, Republic of Korea}\\
\normalsize{$^2$Department of Physics, Pukyong National University, Busan 48513, Republic of Korea}\\
\normalsize{$^3$Department of Physics, Hanyang University, Seoul 04763, Republic of Korea}\\
\normalsize{E-mail: $^{a}$hc2725@gmail.com, $^{b}$moonjippark@hanyang.ac.kr}\\
\normalsize{$^\dagger$Corresponding authors equally contributing to this work.}
\vspace{-0.1cm}}

\date{}

\begin{document} 

\baselineskip24pt

\maketitle

    \begin{sciabstract}
    Universality class of wave chaos emerges in many areas of science, such as molecular dynamics, optics, and network theory. In this work, we generalize the wave chaos theory to cavity lattice systems by discovering the intrinsic coupling of the crystal momentum to the internal cavity dynamics. The \textit{cavity-momentum locking} substitutes the role of the deformed boundary shape in the ordinary single microcavity problem, providing a new platform for the in-situ study of microcavity light dynamics. The transmutation of wave chaos in periodic lattices leads to a phase space reconfiguration that induces a dynamical localization transition. The degenerate scar-mode spinors hybridize and non-trivially localize around regular islands in phase space. In addition, we find that the momentum coupling becomes maximal at the Brillouin zone boundary, so the intercavity chaotic modes coupling and wave confinement are significantly altered. Our work pioneers the study of intertwining wave chaos in periodic systems and provide useful applications in light dynamics control.
    \end{sciabstract}

\clearpage

\section*{Inroduction}

    Light dynamics in optical microcavity~\cite{jan-cao,nockel1997ray,PhysRevLett.120.093902,PhysRevA.98.023851,PhysRevResearch.3.023202,PhysRevA.84.063828} provides the prominent platform to study quantum-classical correspondence, formally known as the field of quantum chaos~\cite{stockmann2000quantum,haake2013quantum,casati2015quantum}. Understanding the chaotic signatures in this transitional regime promotes the future technological applications~\cite{jiang2017chaos,bittner2018suppressing,yang2018fighting,chen2019regular} emerging at the interface between classical and quantum mechanics realms or equivalently ray and wave realms in optical microcavities~\cite{jan-cao,nockel1997ray,harayama1,harayama2,harayama3}. Yet, in a more general sense, the level statistics of the microcavity mirror  universal behaviors observed in various chaotic physical systems, such as Rydberg atoms~\cite{assmann2016quantum,hunter2020rydberg,facon2016sensitive}, ultra-cold atoms~\cite{arnal2020chaos,frisch2015ultracold,frisch2014quantum}, quantum dots~\cite{ponomarenko2008chaotic,albert2011observing,moore1994observation}, and many-body systems~\cite{sa2020complex,pandey2020adiabatic,friedman2019spectral,dahan2022classical}.

    The boundary deformation (BD) in microcavity has been considered the most common approach for phase space engineering to yield the desired optical properties. The main idea is based on the observation that BD sensitively reconfigures the underlying phase space transportation~\cite{PhysRevLett.100.033901,PhysRevLett.100.174102}. Alternative approaches also have been proposed, such as tailoring phase space by defect~\cite{PhysRevLett.127.273902} or including a circular hole inside the cavity~\cite{PhysRevA.73.031802,PhysRevA.79.063804}. Despite these comprehensive efforts, previous approaches still require a holistic device change at the fabrication stage, which remains the main obstacle to further rapid progress. In this regard, developing a new platform on which the light dynamics can be studied by \textit{in-situ} experimental control is genuinely appealing. Here, we propose a two-dimensional periodic lattice structure consisting of multiple chaotic microcavities as a promising breakthrough. We reveal that the external crystal momentum coupling to the internal cavity dynamics can take over the role of the BD by breaking and restoring the inherent symmetry of the cavity. Both the direction and the amplitude of the external momentum can be readily controlled by steering the coherent excitation light sources. As a result, cavity lattice systems promise a hold for feasible control of wave chaos features such as dynamical localization and tunneling~\cite{davis1981quantum,keshavamurthy2011dynamical,PhysRevLett.100.174103,PhysRevLett.113.174101,PhysRevLett.117.144104}.

    In this work, we explore chaotic signatures in periodic lattice dictated by the Bloch theorem for the first time. We study the adiabatic change of the internal cavity states due to the crystal momentum coupling as
    \bea
    \Psi_{\textrm{tot}}(\mathbf{r})=e^{i\mathbf{k}\cdot \mathbf{r}} \psi_{\textrm{Int}}(\mathbf{r})\ , 
    \eea
    where $\Psi_{\textrm{tot}}(\mathbf{r})$ and $\mathbf{k}$ represent the total wave function over the lattice and the crystalline momentum, respectively. Here, $\psi_{\textrm{Int}}(\mathbf{r})$ describes the internal dynamics of the cavity states, which we focus on.
    The coupling between the internal phase space dynamics and the crystal momentum is observed explicitly. In particular, the momentum-induced coupling leads to the hybridizations of the wave function from the scar states to the regular orbit states; namely, we observe the crystal momentum-induced \textit{dynamical localization transition} of the internal dynamics. On the lattice scale, we find that introducing the additional deformation leads to the finite Berry curvature and Hall effect triggered by skew scattering events. Our work firstly promotes applications to topological optical transport utilizing chaotic states. Finally, we discuss the possibility of extending our studies to various lattice systems.
    \begin{figure}[h!t]
   \centering\includegraphics[width=1.\columnwidth]{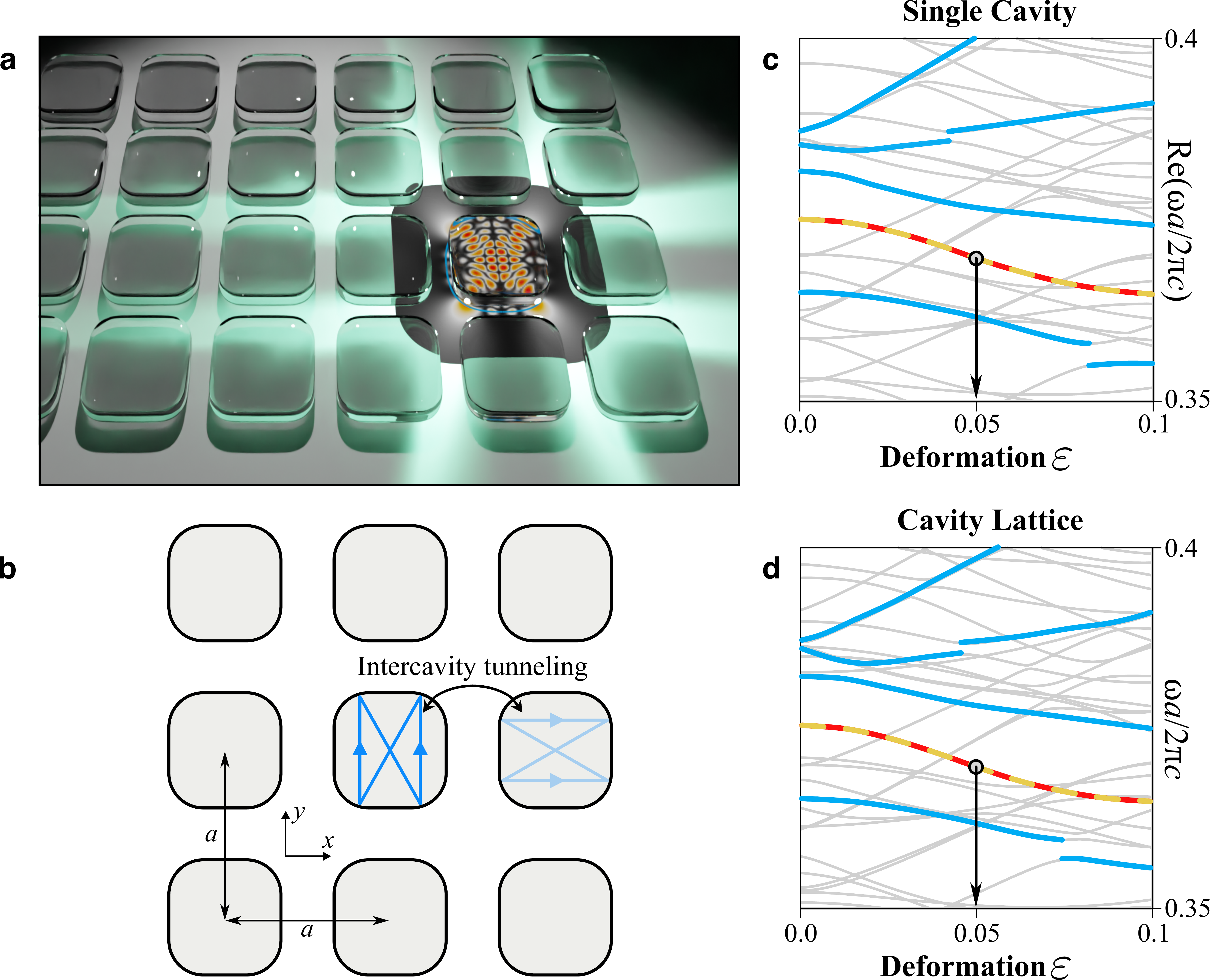}
    \caption{\textbf{a} Conceptual illustration of a photonic crystal consisting of a deformed dielectric microcavities. \textbf{b} Schematic diagram of a square lattice unit-cell with a lattice constant $a$. \textbf{c} Real part resonant frequencies in a single cavity as a function of the cavity deformation $\varepsilon$. \textbf{d} Energy eigenvalues in a cavity array lattice as a function of $\varepsilon$ at zero crystalline momentum $\mathbf{k}=0$ [$\Gamma$ in Fig.~\ref{f4}\textbf{a}]. In \textbf{c} and \textbf{d}, successive Demkov-type couplings 
    for the stable regular modes and the scar modes are highlighted by thick curves. The arrows at $\varepsilon=0.05$ mark the degenerated scar modes (solid and dashed curves) we study.}
    \label{f1}
    \end{figure}


    We consider a square lattice consisting of a single dielectric cavity with a relative refractive index, $n_{\textrm{in}}/n_{\textrm{out}}=10$, inside the cavity [see Fig.~\ref{f1}]. With this high refractive index, we can find energetically well-isolated target modes by suppressing undesired overlaps with other extra modes. However, our result is generally applicable to the broad range of the refractive index. The boundary of each cavity is determined by the four-fold rotational symmetric (C$_4$) deformation, which can be represented as, $r(\theta;\varepsilon)=r_0[1-\varepsilon \cos (4\theta)]$, where $\varepsilon$ denotes the deformation strength. $r_0=R/\sqrt{1+\varepsilon^2/2}$, where $R$ is the radius of the circle when $\varepsilon=0$, is the normalization constant preserving a cavity area under the variation of $\varepsilon$. The Helemholtz equation, $-\nabla^2\vec{\psi}=n(x,y)\frac{\omega}{c}\vec{\psi}$, is solved to obtain resonant frequencies $\omega$ of the transverse-magnetic (TM)-polarized modes [$\vec{\psi}=(0,0,E_z)$], where $c$ is a speed of light. For TM-modes, the waves and their normal derivative are set to be continuous across the cavity boundary. The two lattice vector, where we impose a periodic boundary condition accordingly, is given by $|a_x|=|a_y|=a=2.2R$. To numerically obtain the modes both in a single cavity and in a lattice, we implement the boundary element method~\cite{BEM_wiersig,veble,BEM_sakurai_bss_gen,BEM_isakari_bss_fmm}.

\begin{figure}[h!t]
    \centering\includegraphics[width=1.\textwidth]{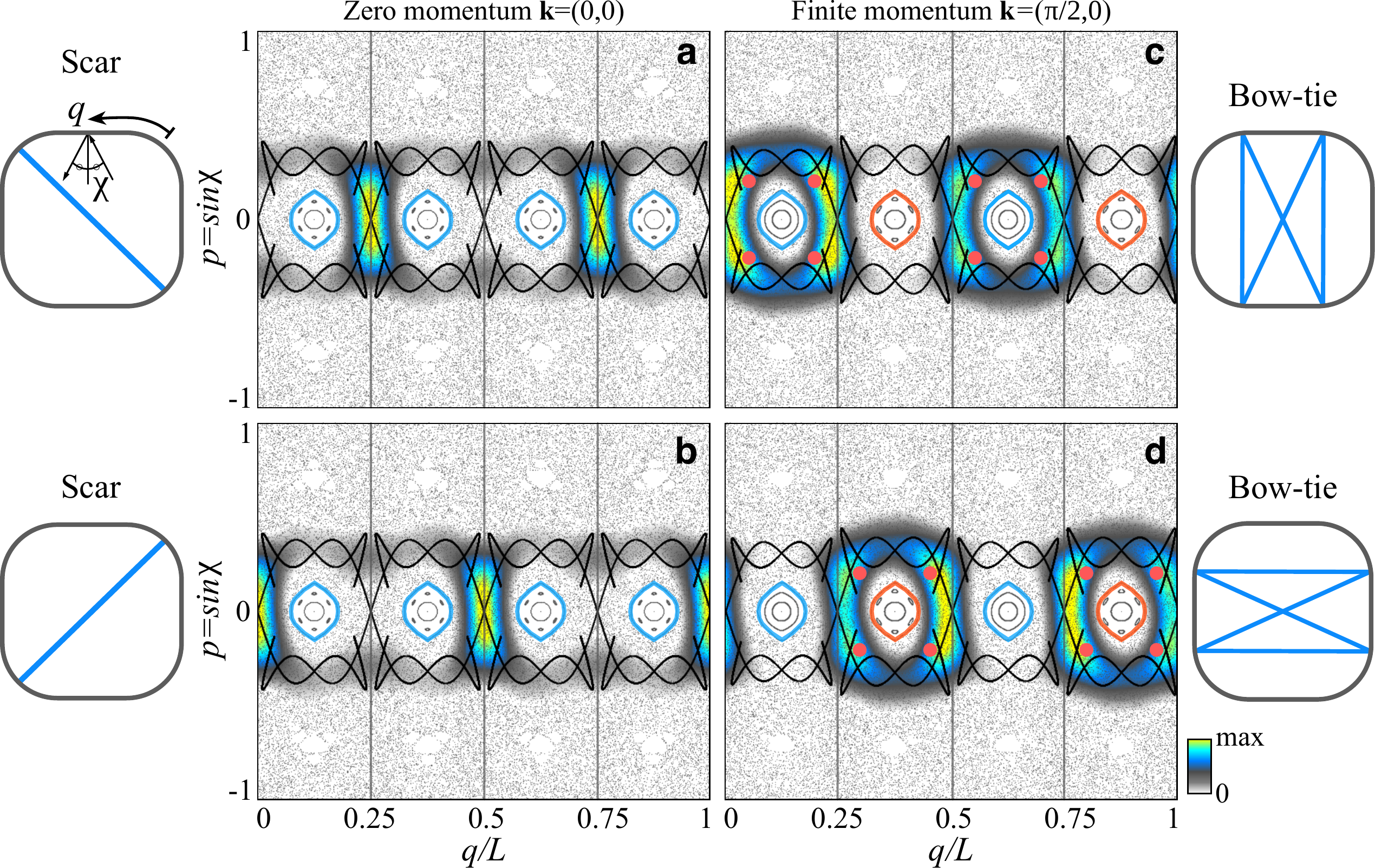}
    \caption{Husimi functions for modes in a cavity lattice superimposed on the ray-dynamical classical phase space (underlying gray dots) in Birkhoff-coordinate $(q/L,p=\sin \chi)$, where $q/L$ is a arclength normalized by a cavity perimeter $L$ and $\chi$ a incident angle of ray. In \textbf{a}-\textbf{d}, the four large closed islands and the crossing points of stable and unstable manifolds (black curves) at $p\sim 0$ correspond period-2 stable and unstable orbits, respectively. The orange dots in \textbf{c} and \textbf{d} represent the phase space points of the bow-tie orbit. \textbf{a} and \textbf{b} show the Husimi distributions for the scar states obtained when $\mathbf{k}=(k_x,k_y)=(0,0)$, while \textbf{c} and \textbf{d} are for the bow-tie orbit states induced by couplings of the two scar states when $\mathbf{k}=(\pi/2,0)$. The leftmost and rightmost panels show corresponding classical orbits in real space. The four vertical lines in \textbf{a}-\textbf{d} indicate a quarter arclength interval $L/4$. When $\mathbf{k}=(0,0)$ (\textbf{a}, \textbf{b}), the phase space (e.g., islands and manifolds) exhibits a $L/4$ translation symmetry, and the Husimi functions in \textbf{a} and \textbf{b} are distributed mutual exclusively. On the other hand, when $\mathbf{k}=(\pi/2,0)$ (\textbf{c}, \textbf{d}), the translation symmetry of the phase space breaks, and the Husimi distributions in \textbf{c} and \textbf{d} have large overlaps around the vertical lines. Detailed correspondences between the Bloch momenta $\mathbf{k}$ and the phase space deformation can be found in Supplementary Materials. The continuous evolution of the Husimi function and the underlying phase space over the range $[\pi/4\le \tan^{-1}(k_y/k_x)\le 9\pi/4]$ for $|\mathbf{k}|=\pi/2$ is shown in the supplementary animations: \quoting{Ani\_supple\_phc\_hus1\_series.mov} and \quoting{Ani\_supple\_phc\_hus2\_series.mov}.}
    \label{f2}
    \end{figure}

    Under a weak BD, invariant tori in the ray-dynamical phase space of the integrable system are destroyed according to the Kolmogorov-Arnold-Moser (KAM) and Poincar\'e-Birkhoff theorems~\cite{tabor1989chaos,strogatz2018nonlinear}. As the deformation increases, the regular and chaotic regions fill the phase space before it becomes fully chaotic and ergodic. Figure~\ref{f2} shows the phase space in the Birkhoff-coordinate $(q/L,p=\sin \chi)\in [0,1]\times[-1,1]$~\cite{birkhoff1927periodic}. Here, $q/L$ is normalized boundary arclength where the ray bounces off ($L$ is a cavity perimeter), and $\chi$ is the incident angle of the ray. The mixed phase space in the figure shows the island structures of regular motions surrounded by a chaotic sea. See Supplementary Materials for details.
    
    \section*{Results}

    Two degenerate scar modes~\cite{heller1984bound}, the non-trivial localization on the unstable fixed points, are observed in the single cavity case for $\omega R /2\pi c \approx 0.375$ at $\epsilon=0.05$. A series of the thick blue curves in Fig.~\ref{f1}\textbf{c} show the successive regular and scar modes that a similar sequence has been analyzed in~\cite{Yi:17}. The scar modes form a pair under the underlying C$_4$-symmetry. In the lattice, with the identical cavity geometry, the spectra at zero momentum exhibit very similar energy eigenstates (Fig.~\ref{f1}\textbf{d}). We observe the pair of scar modes equally for almost the same energy as the single cavity case. Figures~\ref{f2}\textbf{a} and \textbf{b} show the Husimi distributions of the two degenerate scar states, which are localized on top of the distinct unstable fixed points. These fixed points correspond to period-2 (i.e., two bounces for one cycle of a classical orbit) unstable periodic orbits shown in the leftmost panels in Fig.~\ref{f2}. The black curves in Fig.~\ref{f2} show the stable and unstable manifolds of the period-2 unstable orbit, and the unstable fixed points corresponding to this orbit are the crossing points of those manifolds. Note that the Husimi distributions are obtained for the cavity boundary wave, and the inside-incident version was employed~\cite{hentschel2003husimi,kim2018husimi}.

    \begin{figure}[t]
    \centering
    \includegraphics[width=1.\textwidth]{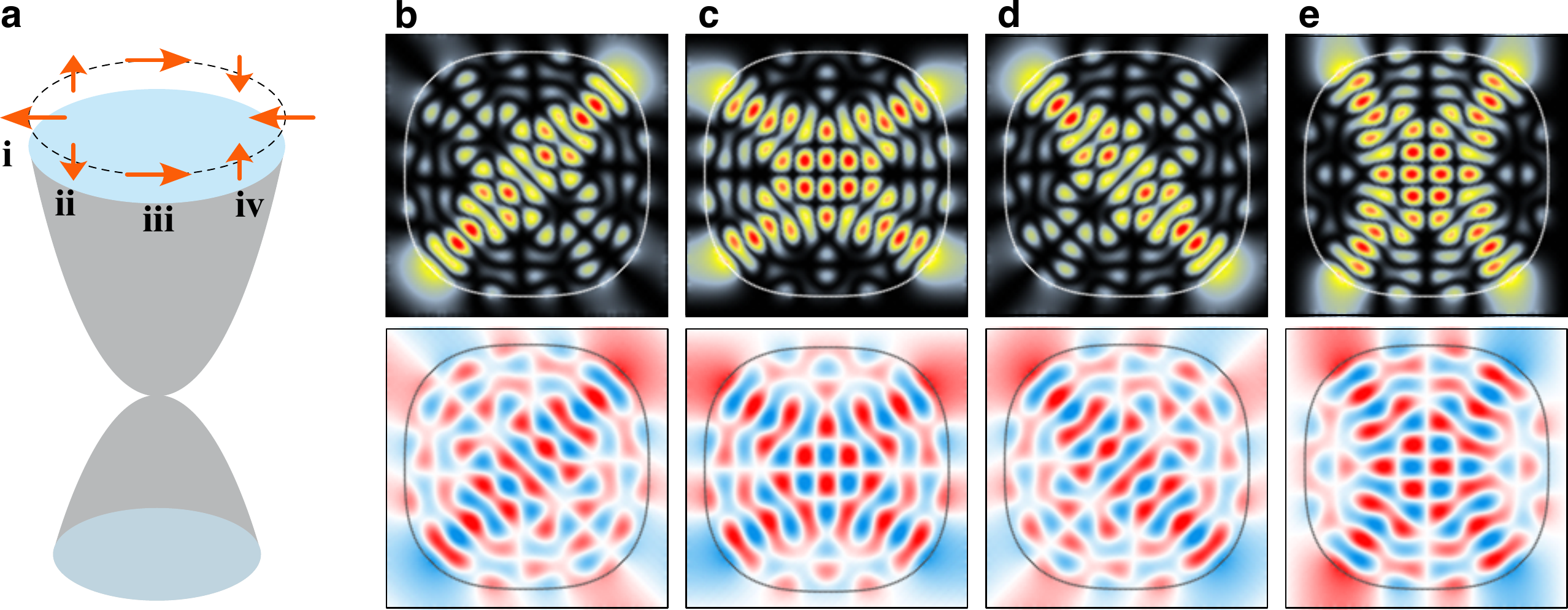}
    \caption{\textbf{a} Schematic illustration of the pseudo-spin evolution along the encircling around the zero crystalline momentum $(k_x,k_y)$=$(0,0)$ ($\Gamma$ in Fig.~\ref{f4}\textbf{a}). \textbf{b}-\textbf{e} The upper and lower panels display the square modulus $|\psi|^2$ and the real part Re($\psi$) of wavefunctions corresponding to, respectively, \textbf{i}-\textbf{iv} in \textbf{a}. The directions of the crystalline momentum for \textbf{b}, \textbf{c}, \textbf{d}, and \textbf{e} are equivalent to the ones of $(k_x,k_y)=(\pi,-\pi)$, $(\pi,0)$, $(\pi,\pi)$, and $(0,\pi)$, respectively.}
    \label{f3}
    \end{figure}

    We now consider the inclusion of the finite crystal momentum $\mathbf{k}=(\textrm{k}_x,\textrm{k}_y)$. In the presence of the non-zero crystal momentum, we observe the degeneracy lifting of the two scar modes. This lifting indicates the state hybridization due to the C$_4$-rotational symmetry breaking of the internal wave function with the non-zero momentum. The hybridized states are found to localize prominently around the period-4 bow-tie orbit (see the rightmost panels in Fig.~\ref{f2}), the satellite orbit of the central period-2 island. Figures~\ref{f2}\textbf{c} and \textbf{d} exemplify the Husimi distributions for these hybridized modes localized on the bow-tie orbits. The mode evolution depending on the crystalline momentum variations in cavity lattices can be equivalently realized in single cavities by applying the additional deformation perturbation. See Supplementary Materials for detailed demonstrations.
    
\begin{figure}[t]
    \centering\includegraphics[width=1.0\textwidth]{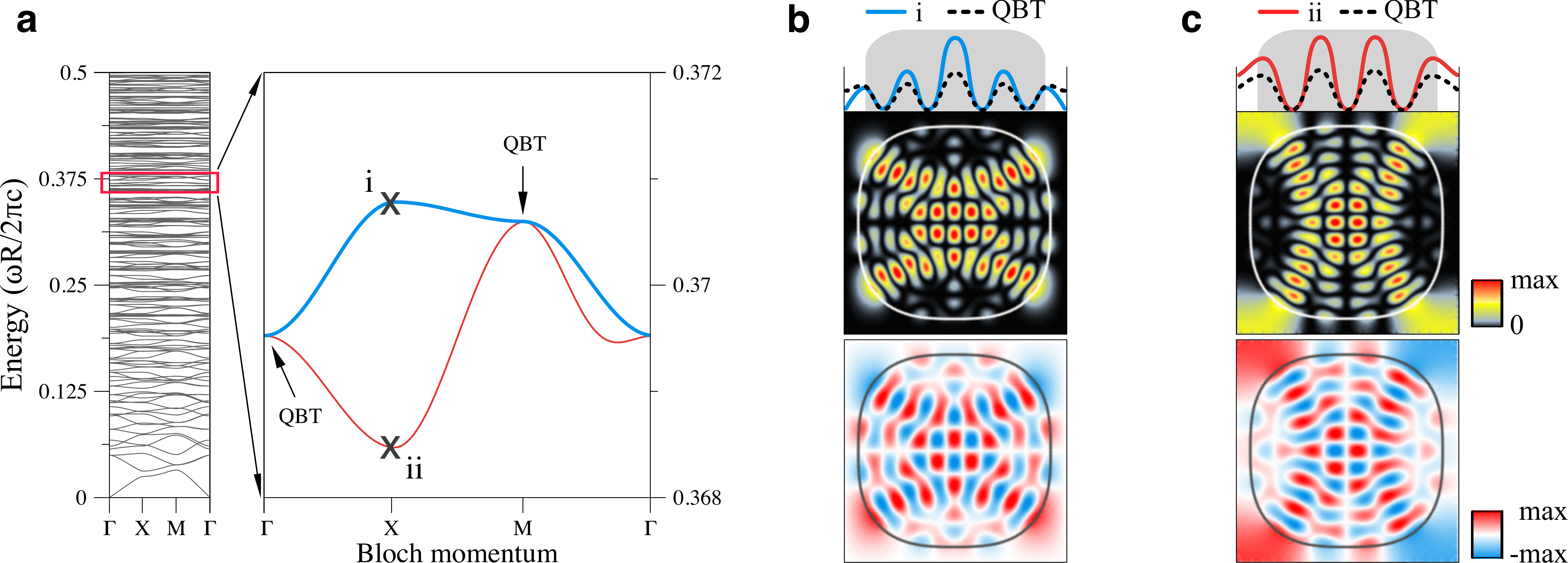}
    \caption{\textbf{a} Energy bands as a function of the crystalline momentum $(k_x,k_y)$ for a fixed cavity deformation $\varepsilon=0.05$. The zoomed bands correspond to the ones indicated by the arrow in Fig.~\ref{f1}\textbf{d}. \textbf{b} Spatial distribution of the even parity mode $(\psi_\textrm{E})$ in a unit-cell for $(k_x,k_y)$=$(\pi,0)$ ($i$ in \textbf{a}). The upper panel depicts the wave projection onto the $x$-axis showing the vanished wave at the unit-cell boundary and dense confinement of wave inside the cavity region (shaded rectangle). \textbf{c} The same as \textbf{b} but for the odd parity mode $(\psi_\textrm{O})$ ($ii$ in \textbf{a}). The upper panel depicts the intense wave distribution outside the cavity, including the unit-cell boundary.} 
    \label{f4}
    \end{figure}

    In addition, Figs.~\ref{f3}\textbf{b}-\textbf{e} show the hybridized states that have strong directional anisotropy in the direction of the crystal momentum. To be specific, the hybridization of the two scar modes is numerically obtained and described by the effective Hamiltonian, given as~\cite{eqderive},
    \bea
    H_{\textrm{SO}}=\frac{\Delta_\textrm{hyb}}{2}(\cos k_x -\cos k_y) \sigma_x + \textrm{V}_0 \sin k_x \sin k_y  \sigma_z\ ,
    \label{e2}
    \eea 
    where $\bm{\sigma}$ represent the Pauli matrices for the scar state degree of freedom. $\Delta_\textrm{hyb}=1.156\times 10^{-3}$ represents the hybridization strength between the two scar modes, while $\textrm{V}_0=0.213\times 10^{-3}$ represents the relative energy shift of the two states, respectively. Since the two scar states are energetically well-decoupled from other states, we can consider the two scar states as the effective spinor $(\ket{\uparrow}$ and $\ket{\downarrow})$. Under this spinor representation, the spinor wind twice around the loop encircling the zero momentum (Fig.~\ref{f3} \textbf{a}). This winding of the spinor gives rise to $2\pi$-Berry phase: $\gamma\equiv i\oint \bra{\psi(\mathbf{k})} \nabla_\mathbf{k} \ket{\psi(\mathbf{k})} d\mathbf{k}=2\pi $, where $\ket{ \psi(\mathbf{k})}$ indicates one of the scar states and $i=\sqrt{-1}$. The degeneracy of the spinor scar states manifests as the topological quadratic band touchings (QBT) protected by C$_4$-symmetry in the momentum space (see Fig.~\ref{f4}\textbf{a}) \cite{PhysRevLett.103.046811}. Here, the mode hybridization due to the crystalline momentum is not restricted in the scar modes and is valid for the general modes that exhibit a degeneracy at time-reversal invariant momenta.

\subsection*{Maximal momentum coupling}
    
    When the momentum becomes maximal near the Brillouin zone (BZ) boundary [$\textrm{X}=(\pi,0)$ and $\textrm{Y}=(0,\pi)$], the wavefunction characteristics in the whole unit-cell domain deviate significantly due to the change in the boundary condition. For instance, at X point, the internal wave function satisfies the anti-periodic boundary condition on the unit-cell boundary: $\psi_\textrm{Int}(x+a,\ y)=-\psi_\textrm{Int}(x,\ y)$ and $\psi_\textrm{Int}(x, y+a)=\psi_\textrm{Int}(x,\ y)$. The linear combination of the two scar states forms two regular states depending on its parity, [$\psi(x, y)=\pm\psi(-x, y)$], due to C$_4$-rotational symmetry. At the X point in BZ, there exist only two possibilities that the resonant energy of the regular state can adiabatically deform. First, even parity regular state, $\psi_\textrm{E}(x,y)$ (Fig.~\ref{f4}\textbf{b}) is not compatible with the anti-periodic boundary condition unless the wave function vanishes at the unit-cell boundary. The vanishing wave function manifests as the flat band with zero group velocity (see the upper band along the X-M interval in Fig.~\ref{f4}\textbf{a}). In addition, since $\psi_\textrm{E}(x,y)$ is confined more densely inside the cavity, where the refractive index is higher, the effective wavelength is reduced, i.e., even parity states form higher energy states (point \quoting{i} in Fig.~\ref{f4}\textbf{a}).

    On the other hand, the odd parity regular state, $\psi_\textrm{O}(x,y)$ (Fig.~\ref{f4}\textbf{c}), shows the compensating intensity enhancement at the unit-cell boundary since the even and odd wave functions must form the complete basis of the two original scar modes. Hence, it results in lengthened effective wavelength (i.e., more waves are outside the cavity, the lower refractive index region) and manifests as the lower energy resonant states (point \quoting{ii} in Fig.~\ref{f4}\textbf{a}). Finally, when the momentum vector reaches the BZ corner [$\textrm{M}=(\pi,\pi)$], the global C$_4$-rotational symmetry is restored. In this case, the hybridization disappears to form the degenerate QBT again, which can be interpreted as a revived dynamical localization of the spinor scar modes.
    \begin{figure}[t!]
    \centering\includegraphics[width=1.\columnwidth]{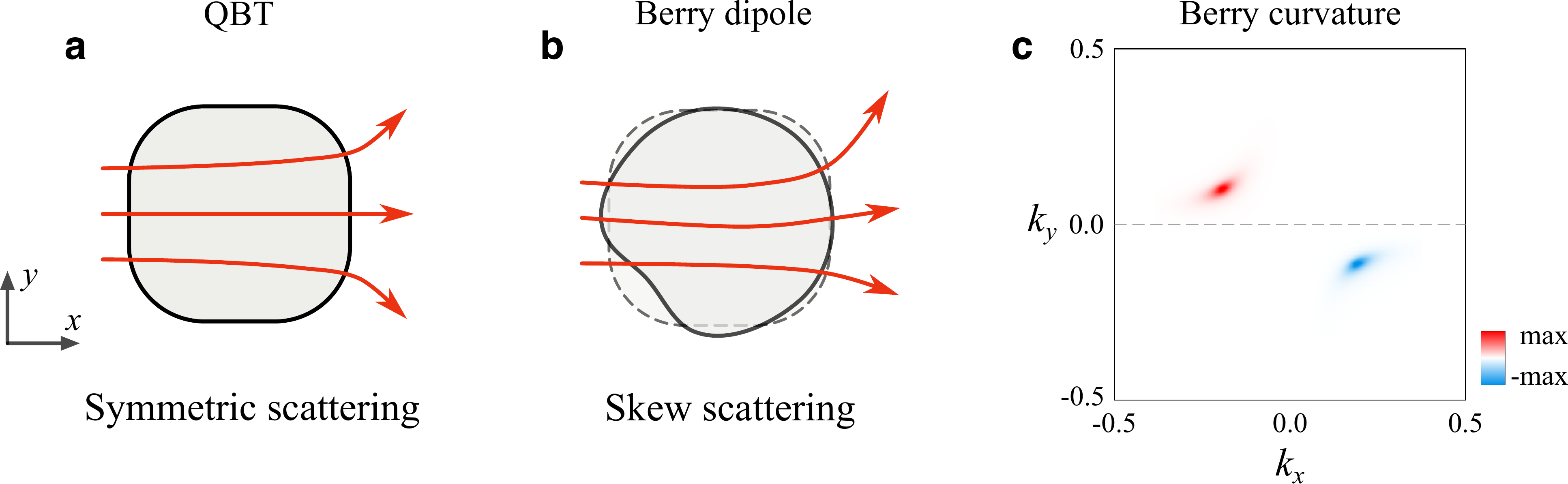}
    \caption{Schematic illustration of BDs \textbf{a} original C$_4$-symmetric boundary with QBT and \textbf{b} C$_4$-breaking boundary. C$_4$-breaking boundary perturbation splits the single QBT into a pair of Dirac cones. Subsequently, C$_2$-breaking boundary perturbation gaps out the Dirac cones and induces the Berry curvature dipole. The Berry curvature dipole induces the skew scattering depending on the incident momentum, $k_{\textrm{inc}}$. \textbf{c} Distribution of non-zero Berry curvature in the momentum space induced by scar states.} 
    \label{f5}
    \end{figure}

\subsection*{Skew boundary scattering in Chaotic states}    
    
    At last, we consider the scar states in the presence of the generic BD perturbation, which is given as, $\delta r (\theta)=\epsilon_1 \cos (N\theta+\phi_0)$. For the even-number oscillating BD [$N\in 2\mathbb{Z}$] [e.g., $(N,\epsilon_1,\phi_0)=(2,0.01,0)$], which preserves C$_2$ symmetry, the rotational symmetry of the cavity is lowered from C$_4$ to C$_2$ symmetry, the topological protection of QBT is lost. Correspondingly, the QBT-carrying $2\pi$-Berry phase can split into a couple of the Dirac linear band crossings, which of each is associated with the $\pi$-Berry phase. On ther other hand, for odd-number oscillating BD [$N\in 2\mathbb{Z}+1$], C$_2$-symmetry breaking induces the skew symmetric scattering of the chaotic modes  [see Fig.~\ref{f5}\textbf{b}], which induces the non-zero Berry curvature $\left\{\Omega_{xy} (\mathbf{k})= i\left[ \braket{\nabla_{k_x} \psi(\mathbf{k})}{\nabla_{k_y} \psi(\mathbf{k})} - (y\leftrightarrow x)\right]\right\} $ in the momentum space. Figure~\ref{f5}\textbf{c} shows the calculated Berry curvature for [$(N,\epsilon_1,\phi_0)=(5,0.01,\frac{\pi}{10})$]. Since the underlying time-reversal symmetry ensures the antisymmetry of the local Berry curvature [$\Omega_{xy} (\mathbf{k})=-\Omega_{xy} (-\mathbf{k})$], the total sum (monopole) of the curvature vanishes. Accordingly, the skew scattering by the BD is characterized by the finite dipole moment of the local Berry curvature~\cite{PhysRevLett.115.216806,RevModPhys.82.1959}:
    \bea
    D_{j}=\sum_\mathbf{k} \partial_{k_j}\Omega_{xy} (\mathbf{k})\ ,
    \eea
    where $j=x,y$.

    Since the optical microcavity array can be excited by the coherent beam source with a well-defined incident momentum, the Berry curvature dipole can be measured through the skew-symmetric beam transport. The semi-classical transport equation~\cite{RevModPhys.82.1539} describes the light dynamics of the Mie regime, and the non-zero Berry dipole manifests as the effective magnetic field  as,
    \bea
    \mathbf{v}_\mathbf{k}=\frac{\partial E_{\mathbf{k}}}{\partial \mathbf{k}}+\dot{\mathbf{k}}\times \Omega_{xy} (\mathbf{k})\hat{z} , \quad \dot{\mathbf{k}}=\nabla n(\mathbf{r})
    \label{Eq:sc}
    \eea
    where $\mathbf{v}_\mathbf{k}$ is the group velocity vector of the scar states having energy $E_\mathbf{k}$, and $n(\mathbf{r})$ is the slowly varying local refractive index. The second term ($\dot{\mathbf{k}}\times \Omega_{xy}$) in Eq. \eqref{Eq:sc} gives rise to skew symmetric light transport that can be captured by control of the incident beam momentum $\mathbf{k}$. The explicit demonstration of the skew symmetric light transportation can be found in Supplementary Materials.

\section*{Discussion}
    
    To conclude, we have studied the wave chaos of the deformed cavity coupled to the external crystalline momentum in a periodic cavity array. The external crystalline momentum now replaces the role of the boundary shape deformation. By controlling the momentum, we observe the dynamical localization transitions. Our work provides a new promising platform, enabling the in-situ study of various wave chaos phenomena. For example, if the additional higher energy and lower energy states are involved, it can induce crystal momentum-induced dynamical tunneling phenomena. This will be the topic of future study.

    In addition, we note that, contrary to the previous studies of the topological photonic crystals (which mainly focus on the rayleigh regime), the governing dynamics of the chaotic state in the Mie regime is described by the semi-classical transport. In this work, we propose, for the first time, the possibility of realizing Berry curvature-induced transport phenomena that utilize the intrinsic wave property of the chaotic states. The crossover between Rayleigh and Mie regime of Berry curvature induced transport would pioneer a new aspect of wave-particle correspondence in the field of wave chaos.

    Our work can be generalized in a few different aspects. Though we have considered the simple square lattice, many other two-dimensional lattices are expected to show qualitatively different behaviors. For instance, away from a simple Bravais lattice, the deformed cavity in Lieb and Kagome lattice is expected to produce a flat band. The intrinsic localization properties would give rise to a stronger coupling within a unit-cell. Further studies on the various lattice systems and wave chaos would also be an intriguing topic for future study.
    
    \section*{Method}

    We solve the Helmholtz wave equation, which is deduced from Maxwell's equations without sources, for optical modes in single cavities and photonic crystals,
\bea
	-\nabla^2\vec{\psi}=n^2(\mathbf{r})\frac{\omega^2}{c^2}\vec{\psi}\ ,
	\label{eq:helm2}
\eea
    where $n(\mathbf{r})$ and $\omega =c k$ are, respectively, the refractive index of the piecewise homogeneous medium and the free-space temporal frequency with vacuum wavenumber $k$ and speed of light $c$. Given the cavity boundary shape, $R(\theta;\varepsilon)=R_0[1-\varepsilon \cos (4\theta)]$, where $R_0=\rho/\sqrt{1+\varepsilon^2/2}$, we set $n(\mathbf{r})=10$ for $|\mathbf{r}|< R(\theta;\varepsilon)$ and $n(\mathbf{r})=1$ otherwise. $\rho$ is the radius of a circle when $\varepsilon=0$. The lattice constant is given as $|a_x|=|a_y|=a=2.2\rho$. We focus on the transverse magnetic [TM; $\vec{\psi} = (0, 0, E_z )$] polarization of modes. The TM-polarized modes fulfill the dielectric boundary condition that $E_z$ and $\vec{\nu}\cdot\nabla E_z$ are continuous across the boundary interface of two different refractive index domains. Here, $\vec{\nu}$ is an outward normal vector of the boundary. Note the transverse-electric [TE; $\vec{\psi} = (0, 0, H_z )$] modes satisfy a different boundary condition: $H_z$ and $1/n^2 \vec{\nu}\cdot\nabla H_z$ are continuous, yet, still the solution of the same wave equation, \eq{eq:helm2}. In addition to the dielectric boundary conditions, we impose a two-dimensional periodic condition at the unit-cell boundary of the two-dimensional periodic lattice. A pure outgoing wave condition at infinity is applied for the modes in the single cavities. To solve \eq{eq:helm2} numerically, we employ the boundary element method (BEM)~\cite{BEM_wiersig,veble} and further implement the block Sakurai--Sugiura method~\cite{BEM_sakurai_bss_gen,BEM_isakari_bss_fmm} to compute the optical modes more efficiently.

\clearpage
\newpage

\bibliographystyle{Science}
\bibliography{TEXT_reference.bib}

\clearpage
\newpage

    \section*{Acknowledgements}
    We acknowledge financial support from the Institute for Basic Science in the Republic of Korea through the project IBS-R024-D1.

    \section*{Author information}
    \paragraph*{Affiliations\\}
    Center for Theoretical Physics of Complex Systems, Institute for Basic Science (IBS), Daejeon, 34126, Republic of Korea\\
    Chang-Hwan Yi, Hee Chul Park, Moon Jip Park

    \section*{Contributions}
    All authors of C.H. Yi, H. C. Park, and M. J., Park contributed this project by proposing the idea, initiating this project, collecting numerical data, analysing numerical data and carrying out mathematical analysis. All authors commented on and wrote the manuscript draft.

    \section*{Corresponding author}
    Correspondence to Hee Chul Park and Moon Jip Park.

    \section*{Ethics declarations}
    \paragraph*{Conflict of interest\\}
    The authors declare no competing interests.

    \section*{Supplementary materials}
    Supplementary Text\\
    Figures S1 to S7\\

\clearpage
\end{document}